\documentclass[twoside,12pt]{article}
\usepackage{epsfig}

\newcommand{\be}{\begin{equation}}
\newcommand{\ee}{\end{equation}}
\newcommand{\bea}{\begin{eqnarray}}
\newcommand{\eea}{\end{eqnarray}}

\begin{document}

\title{ \vspace{1cm} Unitary and analytic model of nucleon EM structure, the
puzzle with JLab proton polarization data and new insight into
proton charge distribution }
\author{C.Adamu\v{s}\v{c}\'{\i}n$^{1},$ S.Dubni\v{c}ka$^{1},$ A.Z.Dubni\v{c}%
kov\'{a}$^{2}$, P.Weisenpacher$^{3}$ \\
\\
$^{1}$Institute of Physics, Slovak Academy of Sciences, Bratislava, Slovak
Republic\\
$^{2}$Department of Theoretical Physics, Comenius University, Bratislava,
Slovak Republic\\
$^{3}$Institute of Informatics, Slovak Academy of Sciences,
Bratislava, Slovak Republic} \maketitle
\begin{abstract}
The Unitary and analytic model of nucleon electromagnetic structure,
describing all existing nucleon form factor data, is briefly reviewed. Then
in the framework of this model the problem of inconsistency of older proton
electric form factor data in space-like region (obtained from $%
e^{-}p\rightarrow e^{-}p$ process by the Rosenbluth technique) with recent
Jefferson Lab data on the ratio $G_{Ep}(t)/G_{Mp}(t)$ (measured in precise
polarization $\overrightarrow{e}^{-}p\rightarrow e^{-}\overrightarrow{p}$
experiment) is suggested to be solved in favour of the latter data which,
however, unlike older data cause an existence of the form factor zero, i.e.
a difraction minimum in $|G_{Ep}(t)|$ around $t=-Q^{2}=-13$GeV$^{2}$. The
new behaviour of $G_{Ep}(t)$ with the zero gives modified proton charge
distribution with enlarged value of the mean square charge radius.
\end{abstract}

\section{Introduction}

The electromagnetic (EM) structure of nucleons is a manifestation of their
quark structure in EM interactions, which can be completely described by
four independent scalar functions of one variable ($t=-Q^{2},$ the squared
four momentum transferred by the virtual photon) called EM form factors
(FF's). The most suitable in extracting experimental information are Sachs
electric $G_{E_{N}}(t)$ and magnetic $G_{M_{N}}(t)$ FF's, which in the Breit
frame give the distribution of charge and magnetization within the nucleon,
respectively.

Since the proton structure discovery in the middle of 50's of the last
century until 2000 abundant and more or less accurate nucleon FF data in the
space-like and time-like regions appeared and as result there are more than
500 experimental points with errors (see Figs 1,2) collected for qualified
analyses.

\begin{figure}[htb]
\centerline{\includegraphics[width=0.45\textwidth]{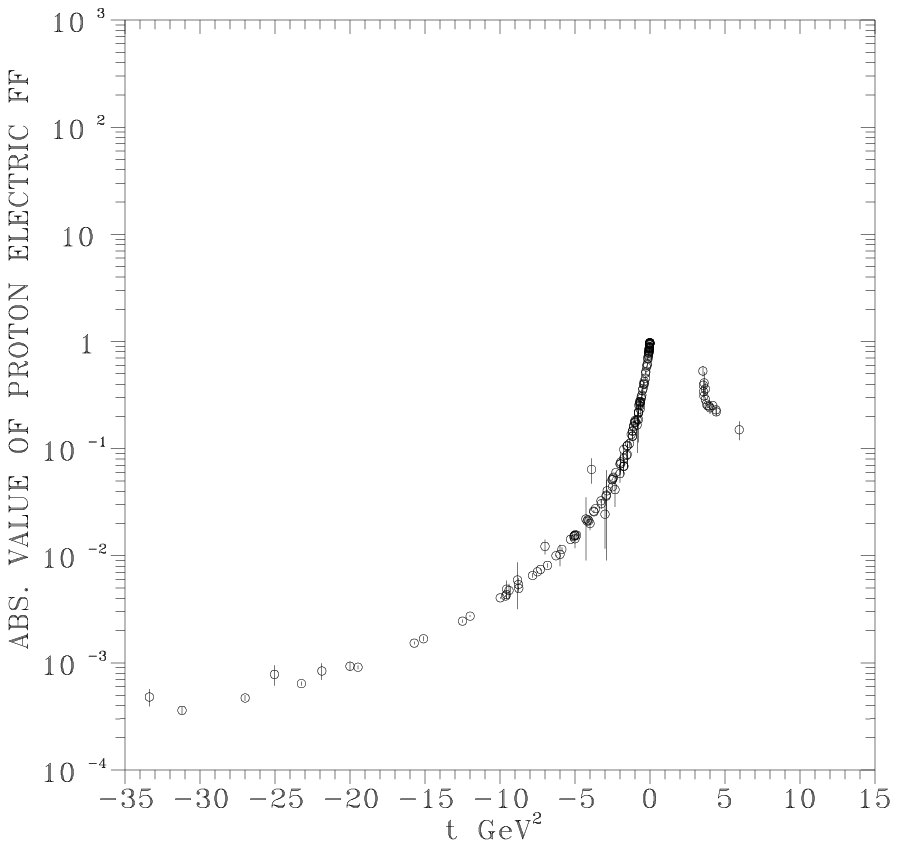}
\qquad
\includegraphics[width=0.45\textwidth]{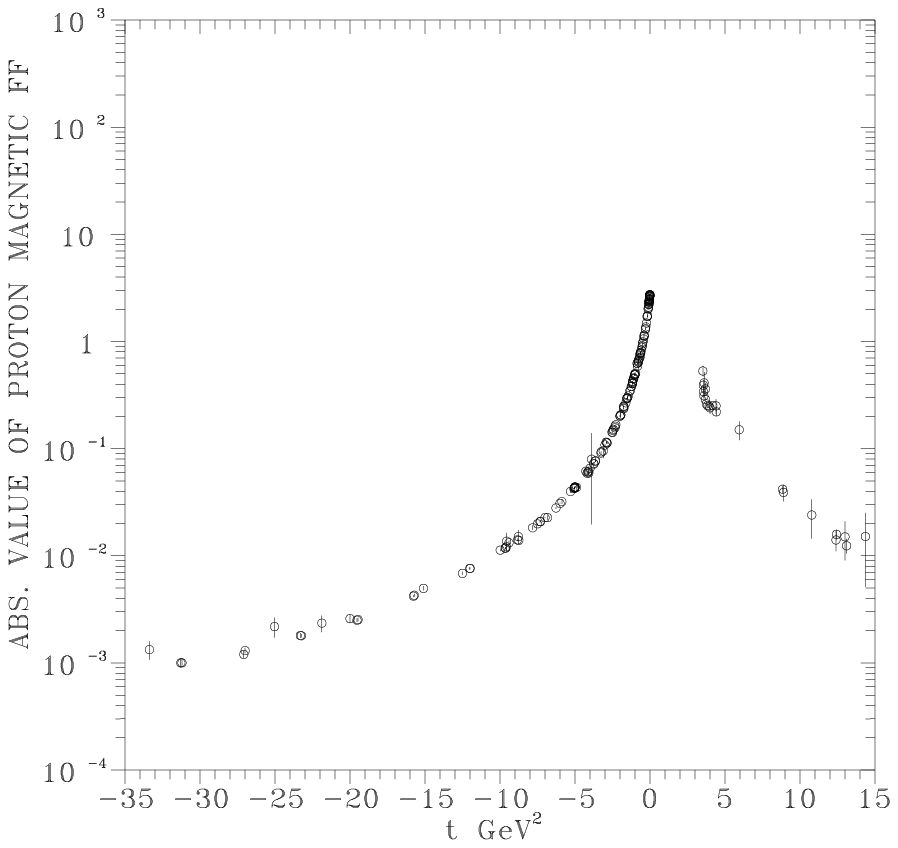}}
\caption{Experimental data on proton electric and magnetic form
factors.}%\label{decay}
\end{figure}

\begin{figure}[htp]
\centerline{\includegraphics[width=0.45\textwidth]{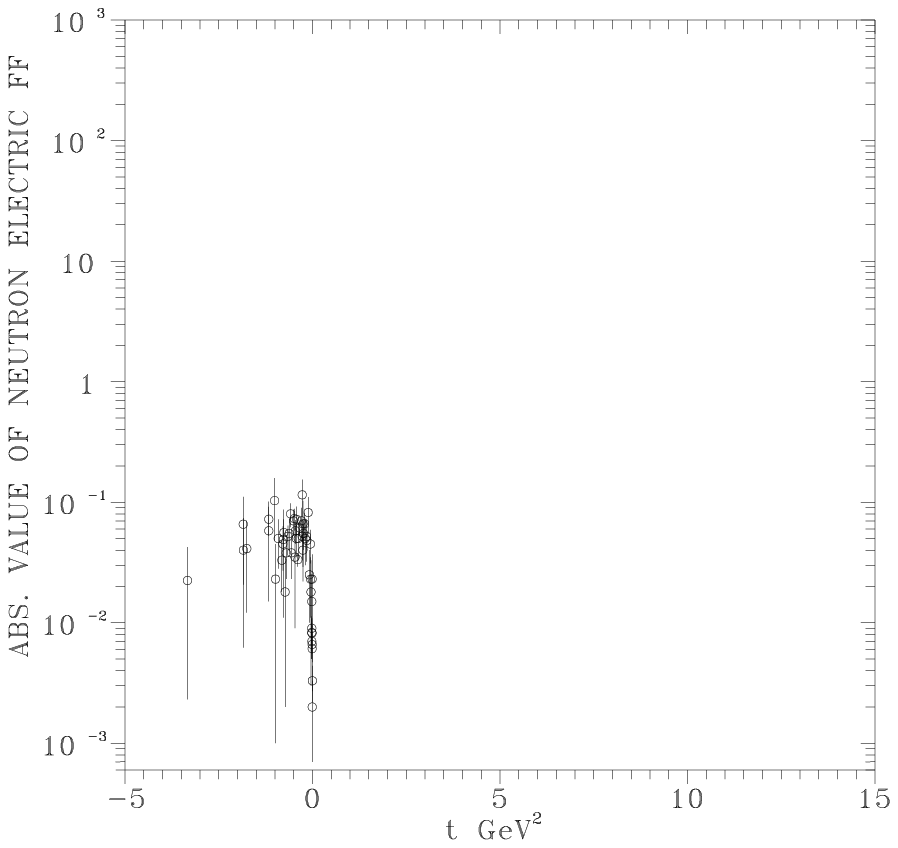}
\qquad
\includegraphics[width=0.45\textwidth]{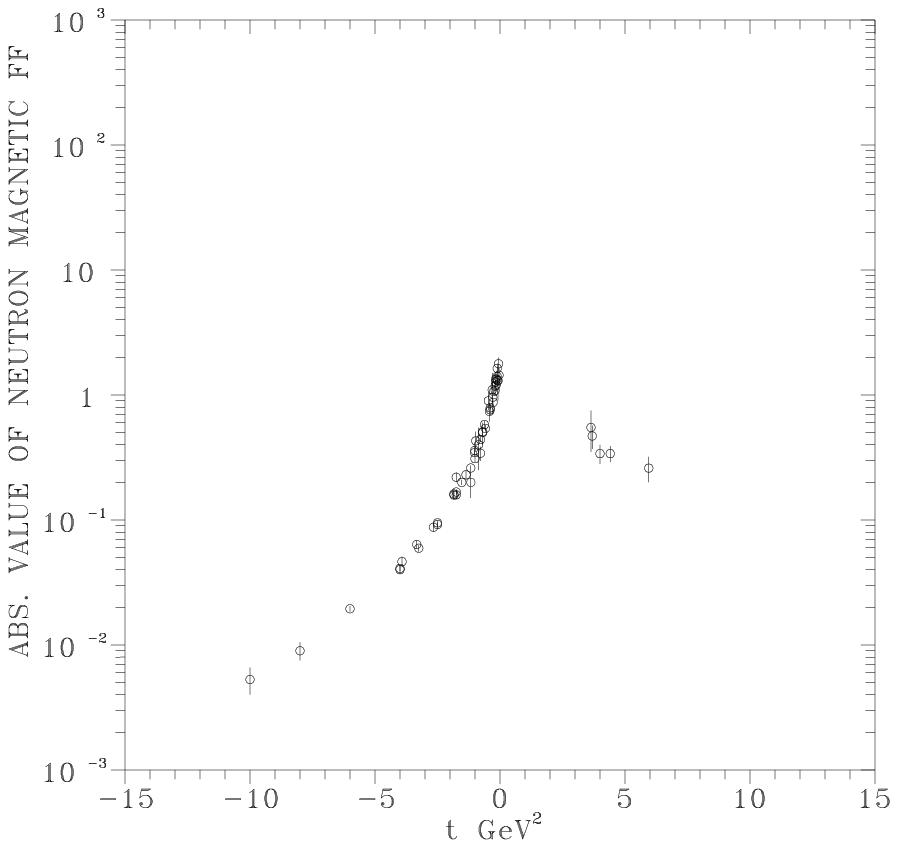}}
\caption{Experimental data on neutron electric and magnetic form
factors.}
%\label{decay}
\end{figure}

Two problems appeared in relation to these so-called older nucleon
FF data, in which proton electric and magnetic FF data in the
space-like region are obtained (mainly in SLAC) from the measured
differential cross sections of the elastic electron-proton
scattering at different electron beam energies
and subsequent utilization of the Rosenbluth technique to separate $%
G_{Ep}(t) $ and $G_{Mp}(t)$.

First, there was no well founded nucleon EM structure model (not even the
dispersion theoretical approach [1-3]) to be successful in a description of
these nucleon FF data in the space-like and time-like region simultaneously.
In particular, the time-like magnetic neutron FF data caused problems.

Second, the recent Jefferson Lab data on $\mu
_{p}G_{Ep}(Q^{2})/G_{Mp}(Q^{2}) $, obtained [4,5] measuring simultaneously
transverse%
\begin{equation}
P_{t}=\frac{h}{I_{0}}(-2)\sqrt{\tau (1+\tau )}G_{Mp}G_{Ep}\tan (\theta /2)
\end{equation}%
and longitudinal%
\begin{equation}
P_{l}=\frac{h(E+E^{\prime })}{I_{0}m_{p}}\sqrt{\tau (1+\tau )}G_{Mp}^{2}\tan
^{2}(\theta /2)
\end{equation}%
components of the recoil proton's polarization in the electron scattering
plane of the polarization transfer process $\overrightarrow{e}%
^{-}p\rightarrow e^{-}\overrightarrow{p}$ ($h$ is the electron beam
helicity, $I_{0}$ is the unpolarized cross-section excluding $\sigma _{Mott}$
and $\tau =Q^{2}/4m_{p}^{2}$) are in a rather strong disagreement with the
data on $G_{Ep}(Q^{2})$ obtained by Rosenbluth technique from differential
cross-section on unpolarized electron scattering on protons.

The first problem was solved successfully in [6] and its solution will be
briefly reviewed in the next Section. Section 3 will be devoted to the
analysis of the second problem, utilizing unitary and analytic nucleon EM
structure model from [6]. In the Section 4 consequences of the analysis of
the Section 3 for the charge distribution within the proton are presented.
The last Section contains a conclusion of the paper.

\section{Unitary and analytic model of nucleon EM structure}

In order to describe data in Figs. 1,2 we construct the model in the
language of isoscalar and isovector parts of the Dirac and Pauli FF's to be
related with Sachs electric and magnetic FF's \ by the expressions%
\begin{eqnarray}
G_{Ep}(t) &=&\left[ F_{1}^{s}(t)+F_{1}^{v}(t)\right] +\frac{t}{4m_{p}^{2}}%
\left[ F_{2}^{s}(t)+F_{2}^{v}(t)\right]  \nonumber \\
G_{Mp}(t) &=&\left[ F_{1}^{s}(t)+F_{1}^{v}(t)\right] +\left[
F_{2}^{s}(t)+F_{2}^{v}(t)\right] \\
G_{En}(t) &=&\left[ F_{1}^{s}(t)-F_{1}^{v}(t)\right] +\frac{t}{4m_{p}^{2}}%
\left[ F_{2}^{s}(t)-F_{2}^{v}(t)\right]  \nonumber \\
G_{Mn}(t) &=&\left[ F_{1}^{s}(t)-F_{1}^{v}(t)\right] +\left[
F_{2}^{s}(t)-F_{2}^{v}(t)\right] ,  \nonumber
\end{eqnarray}%
where then all known nucleon FF properties, like the experimental fact of a
creation of unstable vector-meson resonances in the $e^{+}e^{-}$
annihilation processes into hadrons, the hypothetical analytic properties
(two-cut approximation) of nucleon EM FF's, by means of which just the
contribution of continua are taken into account, the reality condition, the
unitarity condition, normalizations, the asymptotic behaviour of FF's as
predicted by quark model of hadrons are included. It has been manifested to
be sufficient to saturate the isoscalar and isovector Dirac and Pauli FF's
by 5 isoscalars ($\omega ,\phi ,\omega ^{\prime },\omega ^{\prime \prime
},\phi ^{\prime }$) and 5 isovectors ($\rho ,\rho ^{\prime },\rho ^{\prime
\prime },\rho ^{\prime \prime \prime },\rho ^{\prime \prime \prime \prime }$%
), first in the form of standard VMD parametrization, then utilizing the
results of [7] it has been transformed to the form of the VMD
parametrization to be automatically normalized with required asymptotic
behaviour and finally, by using of the special nonlinear transformation and
an introduction of the instability ($\Gamma _{v}\neq 0$) of the vector
mesons under consideration, in the form of ten-resonance unitary and
analytic model of nucleon EM structure as follows%
\begin{eqnarray}
F_{1}^{s}[V(t)] &=&\left( \frac{1-V^{2}}{1-V_{N}^{2}}\right) ^{4}\left\{
\frac{1}{2}H_{\omega ^{\prime \prime }}(V)\cdot L_{\omega ^{\prime }}(V)+%
\left[ H_{\omega ^{\prime \prime }}(V)\cdot L_{\omega }(V)\cdot \frac{%
C_{\omega ^{\prime \prime }}^{1s}-C_{\omega }^{1s}}{C_{\omega ^{\prime
\prime }}^{1s}-C_{\omega ^{\prime }}^{1s}}\right. \right. -  \nonumber \\
&-&L_{\omega ^{\prime }}(V)\cdot L_{\omega }(V)\frac{C_{{\omega^{^{\prime}}}%
}^{1s}-C_{\omega}^{1s}}{C_{\omega^{^{\prime\prime}}}^{1s}-C_{\omega^{^{%
\prime}}}^{1s}}-H_{\omega ^{\prime \prime }}(V)\cdot L_{\omega ^{\prime }}(V)%
\biggr ](f_{\omega {NN}}^{(1)}/f_{\omega })+  \nonumber \\
&+&\left[ H_{\omega ^{\prime \prime }}(V)\cdot L_{\phi }(V)\frac{%
C_{\omega^{^{\prime\prime}}}^{1s}-C_{\phi}^{1s}}{C_{\omega^{^{\prime%
\prime}}}^{1s}-C_{\omega^{^{\prime}}}^{1s}}-L_{\omega ^{\prime }}(V)\cdot
L_{\phi }(V)\frac{C_{\omega^{^{\prime}}}^{1s}-C_{\phi}^{1s}}{%
C_{\omega^{^{\prime\prime}}}^{1s}-C_{\omega^{^{\prime}}}^{1s}}\right. -
\label{d14} \\
&-&H_{\omega ^{\prime \prime }}(V)\cdot L_{\omega ^{\prime }}(V)\biggr ](f_{{%
\phi }{NN}}^{(1)}/f_{{\phi }})-\left[ H_{\phi ^{\prime }}(V)\cdot H_{\omega
^{\prime \prime }}(V)\frac{C_{\phi^{^{\prime}}}^{1s}-C_{\omega^{^{\prime%
\prime}}}^{1s}}{C_{\omega^{^{\prime\prime}}}^{1s}-C_{\omega^{^{\prime}}}^{1s}%
}\right. -  \nonumber \\
&-&H_{\phi ^{\prime }}(V)\cdot L_{\omega ^{\prime }}(V)\frac{%
C_{\phi^{^{\prime}}}^{1s}-C_{\omega^{^{\prime}}}^{1s}}{C_{\omega^{^{\prime%
\prime}}}^{1s}-C_{\omega^{^{\prime}}}^{1s}}+H_{\omega ^{\prime \prime
}}(V)\cdot L_{\omega ^{\prime }}(V)\biggr ](f_{{\phi ^{^{\prime }}}{NN}%
}^{(1)}/f_{{\phi ^{^{\prime }}}})\biggr\}  \nonumber
\end{eqnarray}

\begin{eqnarray}
F_{1}^{v}[W(t)] &=&\left( \frac{1-W^{2}}{1-W_{N}^{2}}\right) ^{4}\left\{
\frac{1}{2}L_{\rho ^{\prime \prime }}(W)\cdot L_{\rho ^{\prime }}(W)+\left[
L_{\rho ^{\prime \prime }}(W)\cdot L_{\rho }(W)\frac{C_{\varrho ^{^{\prime
\prime }}}^{1v}-C_{\varrho }^{1v}}{C_{\varrho ^{^{\prime \prime
}}}^{1v}-C_{\varrho ^{^{\prime }}}^{1v}}-\right. \right.   \nonumber \\
&-&L_{\rho ^{\prime }}(W)\cdot L_{\rho }(W)\frac{C_{\varrho ^{^{\prime
}}}^{1v}-C_{\varrho }^{1v}}{C_{\varrho ^{^{\prime \prime }}}^{1v}-C_{\varrho
^{^{\prime }}}^{1v}}-L_{\rho ^{\prime \prime }}(W)\cdot L_{\rho ^{\prime
}}(W)\biggr](f_{{\varrho }{NN}}^{(1)}/f_{{\varrho }})+  \nonumber \\
&+&\left[ H_{\rho ^{\prime \prime \prime }}(W)\cdot L_{\rho ^{\prime }}(W)%
\frac{C_{\varrho ^{^{\prime \prime \prime }}}^{1v}-C_{\varrho ^{^{\prime
}}}^{1v}}{C_{\varrho ^{^{\prime \prime }}}^{1v}-C_{\varrho ^{^{\prime
}}}^{1v}}\right. -H_{\rho ^{\prime \prime \prime }}(W)\cdot L_{\rho ^{\prime
\prime }}(W)\frac{C_{\varrho ^{^{\prime \prime \prime }}}^{1v}-C_{\varrho
^{^{\prime \prime }}}^{1v}}{C_{\varrho ^{^{\prime \prime }}}^{1v}-C_{\varrho
^{^{\prime }}}^{1v}}- \label{d15} \\
&-&L_{\rho ^{\prime \prime }}(W)\cdot L_{\rho ^{\prime }}(W)\biggr ](f_{{%
\varrho ^{^{\prime \prime \prime }}}{NN}}^{(1)}/f_{{\varrho ^{^{\prime
\prime \prime }}}})-\left[ H_{\rho ^{\prime \prime \prime \prime }}(W)\cdot
L_{\rho ^{\prime \prime }}(W)\frac{C_{\varrho ^{^{\prime \prime \prime
\prime }}}^{1v}-C_{\varrho ^{^{\prime \prime }}}^{1v}}{C_{\varrho ^{^{\prime
\prime }}}^{1v}-C_{\varrho ^{^{\prime }}}^{1v}}\right.   \nonumber \\
&&-H_{\rho ^{\prime \prime \prime \prime }}(W)\cdot L_{\rho ^{\prime }}(W)%
\frac{C_{\varrho ^{^{\prime \prime \prime \prime }}}^{1v}-C_{\varrho
^{^{\prime }}}^{1v}}{C_{\varrho ^{^{\prime \prime }}}^{1v}-C_{\varrho
^{^{\prime }}}^{1v}}+L_{\rho ^{\prime \prime }}(W)\cdot L_{\rho ^{\prime
}}(W)\biggr](f_{{\varrho ^{^{\prime \prime \prime \prime }}}{NN}}^{(1)}/f_{{%
\varrho ^{^{\prime \prime \prime \prime }}}})\biggr\} \nonumber
\end{eqnarray}

\begin{eqnarray}
F_{2}^{s}[U(t)] &=&\left( \frac{1-U^{2}}{1-U_{N}^{2}}\right) ^{6}\left\{
\frac{1}{2}(\mu _{p}+\mu _{n})H_{\omega ^{\prime \prime }}(U)\cdot L_{\omega
^{\prime }}(U)\cdot L_{\omega }(U)\right. +  \nonumber \\
&+&\left[ H_{\omega ^{\prime \prime }}(U)\cdot L_{\phi }(U)\cdot L_{\omega
}(U)\frac{C_{\omega ^{^{\prime \prime }}}^{2s}-C_{\phi }^{2s}}{C_{\omega
^{^{\prime \prime }}}^{2s}-C_{\omega ^{^{\prime }}}^{2s}}.\frac{C_{\phi
}^{2s}-C_{\omega }^{2s}}{C_{\omega ^{^{\prime }}}^{2s}-C_{\omega }^{2s}}%
\right. +  \nonumber \\
&+&H_{\omega ^{\prime \prime }}(U)\cdot L_{\omega ^{\prime }}(U)\cdot
L_{\phi }(U)\frac{C_{\omega ^{^{\prime \prime }}}^{2s}-C_{\phi }^{2s}}{%
C_{\omega ^{^{\prime \prime }}}^{2s}-C_{\omega }^{2s}}.\frac{C_{\omega
^{^{\prime }}}^{2s}-C_{\phi }^{2s}}{C_{\omega ^{^{\prime }}}^{2s}-C_{\omega
}^{2s}}-  \nonumber \\
&-&L_{\omega ^{\prime }}(U)\cdot L_{\phi }(U)\cdot L_{\omega }(U)\frac{%
C_{\omega ^{^{\prime }}}^{2s}-C_{\phi }^{2s}}{C_{\omega ^{^{\prime \prime
}}}^{2s}-C_{\omega ^{^{\prime }}}^{2s}}.\frac{C_{\phi }^{2s}-C_{\omega }^{2s}%
}{C_{\omega ^{^{\prime \prime }}}^{2s}-C_{\omega }^{2s}}-  \label{d16} \\
&&-H_{\omega ^{\prime \prime }}(U)\cdot L_{\omega ^{\prime }}(U)\cdot
L_{\omega }(U)\biggr](f_{{\phi }{NN}}^{(2)}/f_{{\phi }})+ \nonumber \\
&+&\left[ H_{\phi ^{\prime }}(U)\cdot H_{\omega ^{\prime \prime }}(U)\cdot
L_{\omega ^{\prime }}(U)\frac{C_{\phi ^{^{\prime }}}^{2s}-C_{\omega
^{^{\prime \prime }}}^{2s}}{C_{\omega ^{^{\prime \prime }}}^{2s}-C_{\omega
}^{2s}}.\frac{C_{\phi ^{^{\prime }}}^{2s}-C_{\omega ^{^{\prime }}}^{2s}}{%
C_{\omega ^{^{\prime }}}^{2s}-C_{\omega }^{2s}}\right. -  \nonumber \\
&-&H_{\phi ^{\prime }}(U)\cdot H_{\omega ^{\prime \prime }}(U)\cdot
L_{\omega }(U)\frac{C_{\phi ^{^{\prime }}}^{2s}-C_{\omega ^{^{\prime \prime
}}}^{2s}}{C_{\omega ^{^{\prime \prime }}}^{2s}-C_{\omega ^{^{\prime }}}^{2s}}%
.\frac{C_{\phi ^{^{\prime }}}^{2s}-C_{\omega }^{2s}}{C_{\omega ^{^{\prime
}}}^{2s}-C_{\omega }^{2s}}+  \nonumber \\
&+&H_{\phi ^{\prime }}(U)\cdot L_{\omega ^{\prime }}(U)\cdot L_{\omega }(U)%
\frac{C_{\phi ^{^{\prime }}}^{2s}-C_{\omega ^{^{\prime }}}^{2s}}{C_{\omega
^{^{\prime \prime }}}^{2s}-C_{\omega ^{^{\prime }}}^{2s}}.\frac{C_{\phi
^{^{\prime }}}^{2s}-C_{\omega }^{2s}}{C_{\omega ^{^{\prime \prime
}}}^{2s}-C_{\omega }^{2s}}-  \nonumber \\
&-&H_{\omega ^{\prime \prime }}(U)\cdot L_{\omega ^{\prime }}(U)\cdot
L_{\omega }(U)\biggr](f_{{\phi ^{^{\prime }}}{NN}}^{(2)}/f_{{\phi ^{^{\prime
}}}})\biggr\}  \nonumber
\end{eqnarray}

\begin{eqnarray}
F_{2}^{v}[X(t)] &=&\left( \frac{1-X^{2}}{1-X_{N}^{2}}\right) ^{6}\left\{
\frac{1}{2}(\mu _{p}-\mu _{n})L_{\rho ^{\prime \prime }}(X)\cdot L_{\rho
^{\prime }}(X)\cdot L_{\rho }(X)\right. +  \nonumber \\
&+&\left[ H_{\rho ^{\prime \prime \prime }}(X)\cdot L_{\rho ^{\prime
}}(X)\cdot L_{\rho }(X)\frac{C_{\varrho^{^{\prime\prime\prime}}}^{2v}-C_{%
\varrho^{^{\prime}}}^{2v}}{C_{\varrho^{^{\prime\prime}}}^{2v}-C_{\varrho^{^{%
\prime}}}^{2v}}.\frac{C_{\varrho^{^{\prime\prime\prime}}}^{2v}-C_{%
\varrho}^{2v}}{C_{\varrho^{^{\prime\prime}}}^{2v}-C_{\varrho}^{2v}}\right. -
\nonumber \\
&-&H_{\rho ^{\prime \prime \prime }}(X)\cdot L_{\rho ^{\prime \prime
}}(X)\cdot L_{\rho }(X)\frac{C_{\varrho^{^{\prime\prime\prime}}}^{2v}-C_{%
\varrho^{^{\prime\prime}}}^{2v}}{C_{\varrho^{^{\prime\prime}}}^{2v}-C_{%
\varrho^{^{\prime}}}^{2v}}.\frac{C_{\varrho^{^{\prime\prime%
\prime}}}^{2v}-C_{\varrho}^{2v}}{C_{\varrho^{^{\prime}}}^{2v}-C_{%
\varrho}^{2v}}+  \nonumber \\
&+&H_{\rho ^{\prime \prime \prime }}(X)\cdot L_{\rho ^{\prime \prime
}}(X)\cdot L_{\rho ^{\prime }}(X)\frac{C_{\varrho^{^{\prime\prime%
\prime}}}^{2v}-C_{\varrho^{^{\prime\prime}}}^{2v}}{C_{\varrho^{^{\prime%
\prime}}}^{2v}-C_{\varrho}^{2v}}.\frac{C_{\varrho^{^{\prime\prime%
\prime}}}^{2v}-C_{\varrho^{^{\prime}}}^{2v}}{C_{\varrho^{^{%
\prime}}}^{2v}-C_{\varrho}^{2v}}-  \nonumber \\
&-&L_{\rho ^{\prime \prime }}(X)\cdot L_{\rho ^{\prime }}(X)\cdot L_{\rho
}(X)\biggr ](f_{{\varrho ^{^{\prime \prime \prime }}}{NN}}^{(2)}/f_{{\varrho
^{^{\prime \prime \prime }}}})+  \label{d17} \\
&+&\left[ H_{\rho ^{\prime \prime \prime \prime }}(X)\cdot L_{\rho ^{\prime
}}(X)\cdot L_{\rho }(X)\frac{C_{\varrho^{^{\prime\prime\prime%
\prime}}}^{2v}-C_{\varrho^{^{\prime}}}^{2v}}{C_{\varrho^{^{\prime%
\prime}}}^{2v}-C_{\varrho^{^{\prime}}}^{2v}}.\frac{C_{\varrho^{^{\prime%
\prime\prime\prime}}}^{2v}-C_{\varrho}^{2v}}{C_{\varrho^{^{\prime%
\prime}}}^{2v}-C_{\varrho}^{2v}}\right. -  \nonumber \\
&-&H_{\rho ^{\prime \prime \prime \prime }}(X)\cdot L_{\rho ^{\prime \prime
}}(X)\cdot L_{\rho }(X)\frac{C_{\varrho^{^{\prime\prime\prime%
\prime}}}^{2v}-C_{\varrho^{^{\prime\prime}}}^{2v}}{C_{\varrho^{^{\prime%
\prime}}}^{2v}-C_{\varrho^{^{\prime}}}^{2v}}.\frac{C_{\varrho^{^{\prime%
\prime\prime\prime}}}^{2v}-C_{\varrho}^{2v}}{C_{\varrho^{^{%
\prime}}}^{2v}-C_{\varrho}^{2v}}+  \nonumber \\
&+&H_{\rho ^{\prime \prime \prime \prime }}(X)\cdot L_{\rho ^{\prime \prime
}}(X)\cdot L_{\rho ^{\prime }}(X)\frac{C_{\varrho^{^{\prime\prime\prime%
\prime}}}^{2v}-C_{\varrho^{^{\prime\prime}}}^{2v}}{C_{\varrho^{^{\prime%
\prime}}}^{2v}-C_{\varrho}^{2v}}.\frac{C_{\varrho^{^{\prime\prime\prime%
\prime}}}^{2v}-C_{\varrho^{^{\prime}}}^{2v}}{C_{\varrho^{^{%
\prime}}}^{2v}-C_{\varrho}^{2v}}-  \nonumber \\
&-&L_{\rho ^{\prime \prime }}(X)\cdot L_{\rho ^{\prime }}(X)\cdot L_{\rho
}(X)\biggr ](f_{{\varrho ^{^{\prime \prime \prime \prime }}}{NN}}^{(2)}/f_{{%
\varrho ^{^{\prime \prime \prime \prime }}}})\biggr\}  \nonumber
\end{eqnarray}%
where
\begin{eqnarray}
&&L_{r}(V)=\frac{(V_{N}-V_{r})(V_{N}-V_{r}^{\ast
})(V_{N}-1/V_{r})(V_{N}-1/V_{r}^{\ast })}{(V-V_{r})(V-V_{r}^{\ast
})(V-1/V_{r})(V-1/V_{r}^{\ast })};  \nonumber \\
&&C_{r}^{1s}=\frac{(V_{N}-V_{r})(V_{N}-V_{r}^{\ast
})(V_{N}-1/V_{r})(V_{N}-1/V_{r}^{\ast })}{-(V-1/V_{r})(V-1/V_{r}^{\ast })}%
;\qquad r=\omega ,\phi ,\omega ^{^{\prime }},  \nonumber \\
&&H_{l}(V)=\frac{(V_{N}-V_{l})(V_{N}-V_{l}^{\ast
})(V_{N}+V_{l})(V_{N}+V_{l}^{\ast })}{(V-V_{l})(V-V_{l}^{\ast
})(V+V_{l})(V+V_{l}^{\ast })};  \nonumber \\
&&C_{l}^{1s}=\frac{(V_{N}-V_{l})(V_{N}-V_{l}^{\ast
})(V_{N}+V_{l})(V_{N}+V_{l}^{\ast })}{-(V_{l}-1/V_{l})(V_{l}^{\ast
}-1/V_{l}^{\ast })};\qquad l=\omega ^{^{\prime \prime }},\phi ^{^{\prime }}
\nonumber \\
&&L_{k}(W)=\frac{(W_{N}-W_{k})(W_{N}-W_{k}^{\ast
})(W_{N}-1/W_{k})(W_{N}-1/W_{k}^{\ast })}{(W-W_{k})(W-W_{k}^{\ast
})(W-1/W_{k})(W-1/W_{k}^{\ast })};  \nonumber \\
&&C_{k}^{1v}=\frac{(W_{N}-W_{k})(W_{N}-W_{k}^{\ast
})(W_{N}-1/W_{k})(W_{N}-1/W_{k}^{\ast })}{-(W_{k}-1/W_{k})(W_{k}^{\ast
}-1/W_{k}^{\ast })};\qquad k=\rho ,\rho ^{^{\prime }},\rho ^{^{\prime \prime
}},  \nonumber \\
&&H_{n}(W)=\frac{(W_{N}-W_{n})(W_{N}-W_{n}^{\ast
})(W_{N}+W_{n})(W_{N}+W_{n}^{\ast })}{(W-W_{n})(W-W_{n}^{\ast
})(W+W_{n})(W+W_{n}^{\ast })};  \nonumber \\
&&C_{n}^{1v}=\frac{(W_{N}-W_{n})(W_{N}-W_{n}^{\ast
})(W_{N}+W_{n})(W_{N}+W_{n}^{\ast })}{-(W_{n}-1/W_{n})(W_{n}^{\ast
}-1/W_{n}^{\ast })};\qquad n=\rho ^{^{\prime \prime \prime }},\rho
^{^{\prime \prime \prime \prime }}  \nonumber \\
&&L_{r}(U)=\frac{(U_{N}-U_{r})(U_{N}-U_{r}^{\ast
})(U_{N}-1/U_{r})(U_{N}-1/U_{r}^{\ast })}{(U-U_{r})(U-U_{r}^{\ast
})(U-1/U_{r})(U-1/U_{r}^{\ast })};  \nonumber \\
&&C_{r}^{2s}=\frac{(U_{N}-U_{r})(U_{N}-U_{r}^{\ast
})(U_{N}-1/U_{r})(U_{N}-1/U_{r}^{\ast })}{-(U-1/U_{r})(U-1/U_{r}^{\ast })}%
;\qquad r=\omega ,\phi ,\omega ^{^{\prime }},  \nonumber \\
&&H_{l}(U)=\frac{(U_{N}-U_{l})(U_{N}-U_{l}^{\ast
})(U_{N}+U_{l})(U_{N}+U_{l}^{\ast })}{(U-U_{l})(U-U_{l}^{\ast
})(U+U_{l})(U+U_{l}^{\ast })};  \nonumber \\
&&C_{l}^{2s}=\frac{(U_{N}-U_{l})(U_{N}-U_{l}^{\ast
})(U_{N}+U_{l})(U_{N}+U_{l}^{\ast })}{-(U_{l}-1/U_{l})(U_{l}^{\ast
}-1/U_{l}^{\ast })};\qquad l=\omega ^{^{\prime \prime }},\phi ^{^{\prime }}
\nonumber \\
&&L_{k}(X)=\frac{(X_{N}-X_{k})(X_{N}-X_{k}^{\ast
})(X_{N}-1/X_{k})(X_{N}-1/X_{k}^{\ast })}{(X-X_{k})(X-X_{k}^{\ast
})(X-1/X_{k})(X-1/X_{k}^{\ast })};  \nonumber \\
&&C_{k}^{2v}=\frac{(X_{N}-X_{k})(X_{N}-X_{k}^{\ast
})(X_{N}-1/X_{k})(X_{N}-1/X_{k}^{\ast })}{-(X_{k}-1/X_{k})(X_{k}^{\ast
}-1/X_{k}^{\ast })};\qquad k=\rho ,\rho ^{^{\prime }},\rho ^{^{\prime \prime
}},  \nonumber \\
&&H_{n}(X)=\frac{(X_{N}-X_{n})(X_{N}-X_{n}^{\ast
})(X_{N}+X_{n})(X_{N}+X_{n}^{\ast })}{(X-X_{n})(X-X_{n}^{\ast
})(X+X_{n})(X+X_{n}^{\ast })};  \nonumber \\
&&C_{n}^{2v}=\frac{(X_{N}-X_{n})(X_{N}-X_{n}^{\ast
})(X_{N}+X_{n})(X_{N}+X_{n}^{\ast })}{-(X_{n}-1/X_{n})(X_{n}^{\ast
}-1/X_{n}^{\ast })};\qquad n=\rho ^{^{\prime \prime \prime }},\rho
^{^{\prime \prime \prime \prime }}  \nonumber \\
&&  \nonumber
\end{eqnarray}

and $V(t)$ (similarly $W(t),U(t)$ and $X(t)$) takes the form%
\begin{equation}
V(t)=i\frac{\sqrt{\left( \frac{t_{in}^{1s}-t_{0}^{s}}{t_{0}^{s}}\right)
^{1/2}+\left( \frac{t-t_{0}^{s}}{t_{0}^{s}}\right) ^{1/2}}-\sqrt{\left(
\frac{t_{in}^{1s}-t_{0}^{s}}{t_{0}^{s}}\right) ^{1/2}-\left( \frac{%
t-t_{0}^{s}}{t_{0}^{s}}\right) ^{1/2}}}{\sqrt{\left( \frac{%
t_{in}^{1s}-t_{0}^{s}}{t_{0}^{s}}\right) ^{1/2}+\left( \frac{t-t_{0}^{s}}{%
t_{0}^{s}}\right) ^{1/2}}+\sqrt{\left( \frac{t_{in}^{1s}-t_{0}^{s}}{t_{0}^{s}%
}\right) ^{1/2}-\left( \frac{t-t_{0}^{s}}{t_{0}^{s}}\right) ^{1/2}}}
\end{equation}%
with $t_{0}^{s}=9m_{\pi }^{2},t_{0}^{v}=4m_{\pi
}^{2},t_{in}^{1v}=t_{in}^{2v}=4m_{N}^{2}$ and $t_{in}^{1s},t_{in}^{2s}$
effective square root branch points, which however can not be fixed at
two-nucleon treshold as in the isoscalar case there is a remarkable
contribution of $K\overline{K}$ intermediate state in the unitarity
condition.

Every FF, (4)-(7), represents always one analytic function for $-\infty
<t<\infty $ to be defined on the four-sheeted Riemann surface, providing in
such a way a very effective framework for a compatible superposition of
complex conjugate vector-meson pole pairs on unphysical sheets and continua
contributions in nucleon EM FF's.

If the parameters of $\rho ,\omega ,\phi ,\rho ^{\prime },\omega ^{\prime
},\phi ^{\prime },\rho ^{\prime \prime },\omega ^{\prime \prime }$ are taken
from Review of Particle Properties [8], the parameters of $\rho ^{\prime
\prime \prime }$ from [9] and the parameters of $\rho ^{\prime \prime \prime
\prime }$ are free we are left in $F_{1}^{s}[V(t)]$ with 4 free parameters,
in $F_{1}^{v}[W(t)]$ with 3 free parameters, in $F_{2}^{s}[U(t)]$ with 3
free parameters and in $F_{2}^{v}[X(t)]$ with 4 free parameters. All of them
are determined in comparison of the model with the collected 512
experimental points on the nucleon EM FF's in space-like and time-like
region simultaneously and the fourth excited state of $\rho $-meson with the
slightly lower mass $m_{\rho ^{\prime \prime \prime \prime }}=1455\pm 53$MeV
than in [10] and $\Gamma _{\rho ^{\prime \prime \prime \prime }}=728\pm 42$%
MeV is determined. The quality of the achieved descriptions is graphically
presented in Figs.3,4 by dashed lines.

\begin{figure}[htb]
\centerline{\includegraphics[width=0.45\textwidth]{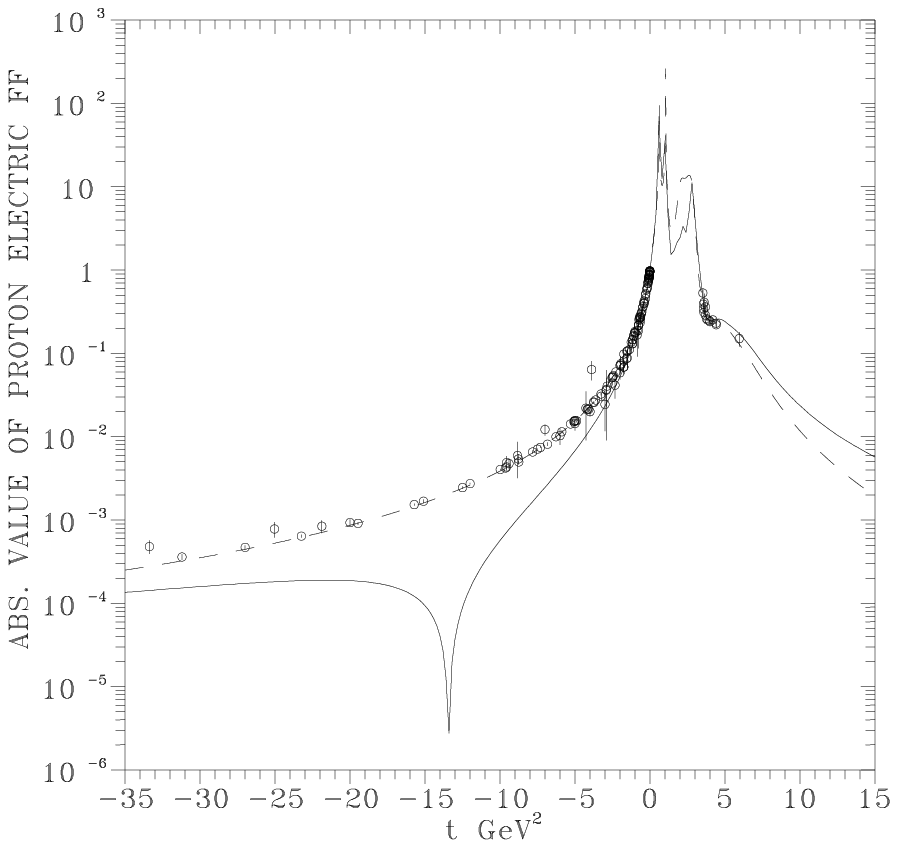}
\qquad
\includegraphics[width=0.45\textwidth]{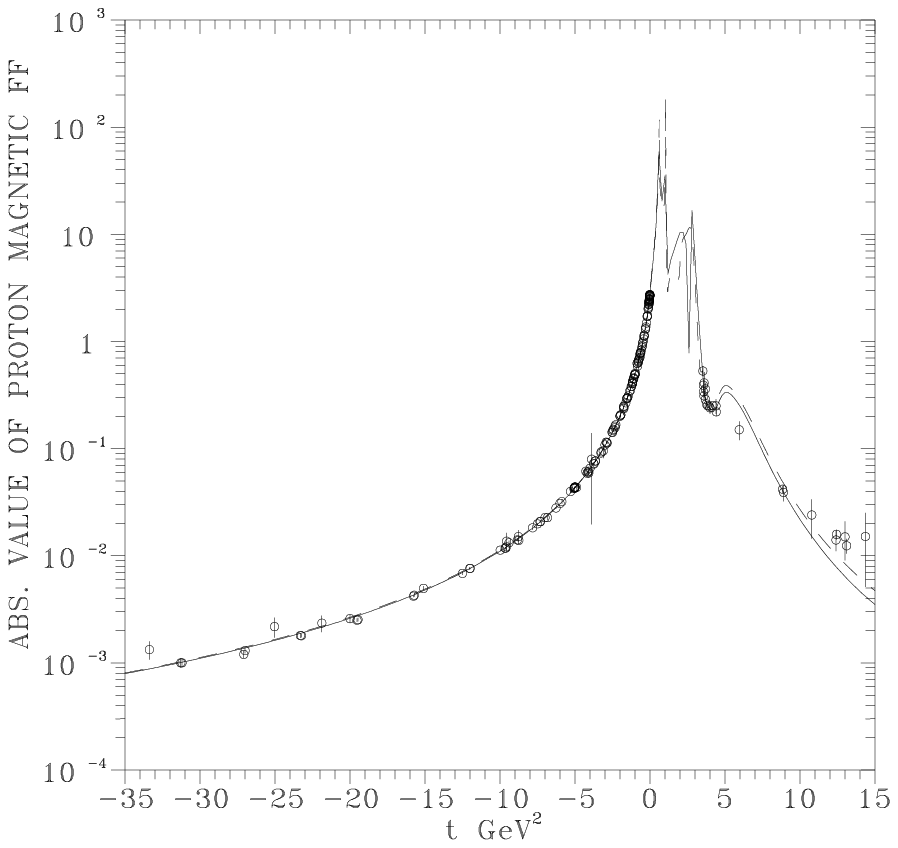}}
\caption{Theoretical behavior of proton electric and magnetic form
factors.}
%\label{decay}
\end{figure}

\begin{figure}[htb]
\centerline{\includegraphics[width=0.45\textwidth]{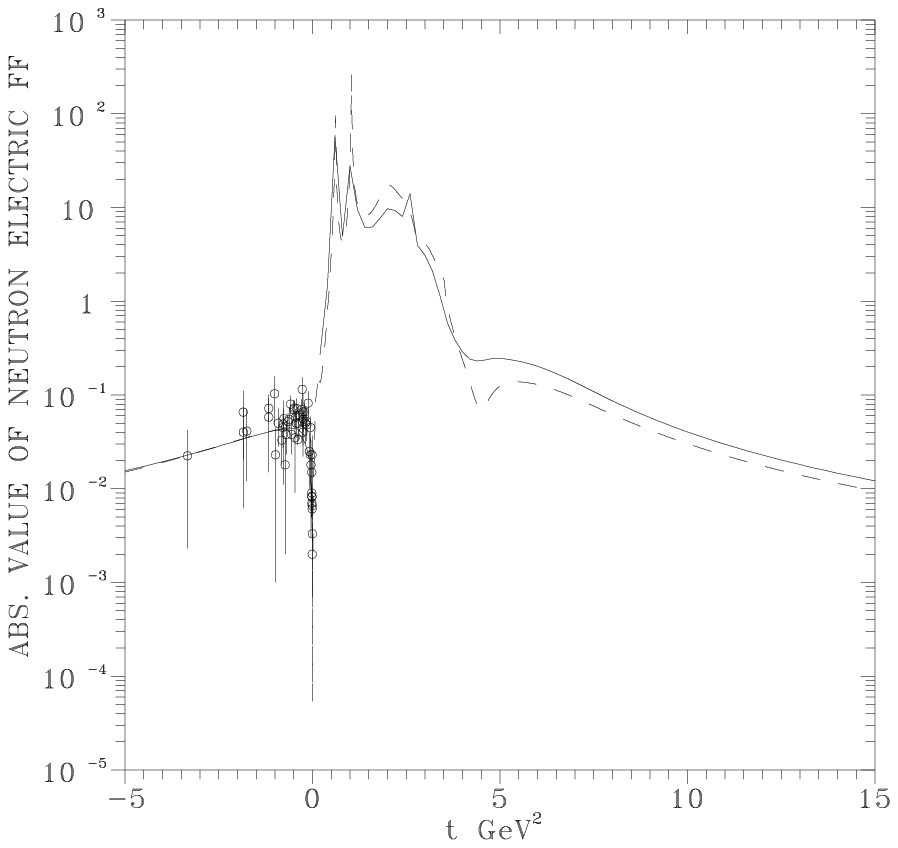}
\qquad
\includegraphics[width=0.45\textwidth]{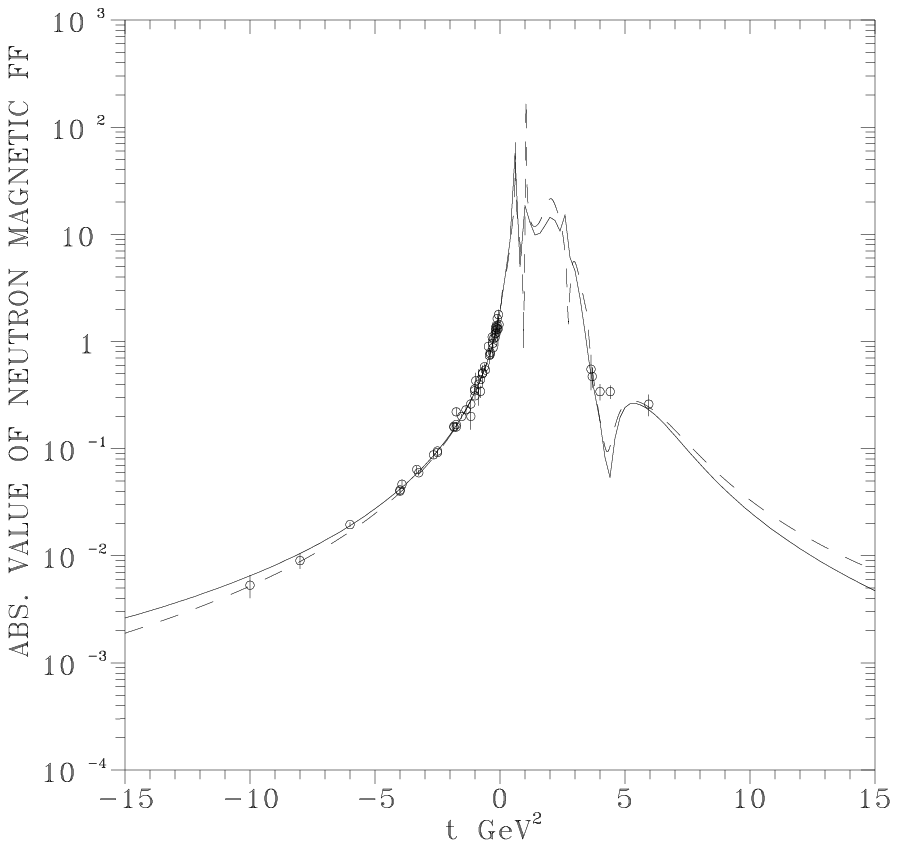}}
\caption{Theoretical behavior of neutron electric and magnetic
form factors.}
%\label{decay}
\end{figure}

\section{JLab polarization data contradicting older SLAC $G_{Ep}(Q^{2})$
behavior and the analysis of the puzzle}

As we have mentioned at the Introduction there are new Jefferson Lab
polarization data [4,5] on $\mu _{p}G_{Ep}(Q^{2})/G_{Mp}(Q^{2})$ for 0.49GeV$%
^{2}\leq Q^{2}\leq 5.54$GeV$^{2}$ (see Fig.5) to be in rather strong
disagreement with the older separate data on $G_{Ep}(Q^{2})$ and $%
G_{Mp}(Q^{2})$ obtained in unpolarized elastic scattering of electrons on
protons by the Rosenbluth technique. Really, whereas the older data on $%
G_{Ep}(Q^{2})$ and $G_{Mp}(Q^{2})$ follow (at least at the region of JLab
data) a dipole behavior and their ratio with increased $Q^{2}$ is almost
constant (see dotted line in Fig.5), the new precise JLab polarization data
on $\mu _{p}G_{Ep}(Q^{2})/G_{Mp}(Q^{2})$ indicate, that $G_{Ep}(Q^{2})$
behavior is somewhere between dipole and tripole one and to some extent it
exhibits a violation of the pQCD prediction for the FF asymptotic behavior
at the measured region.

\begin{figure}[htb]
\centerline{\includegraphics[width=0.45\textwidth]{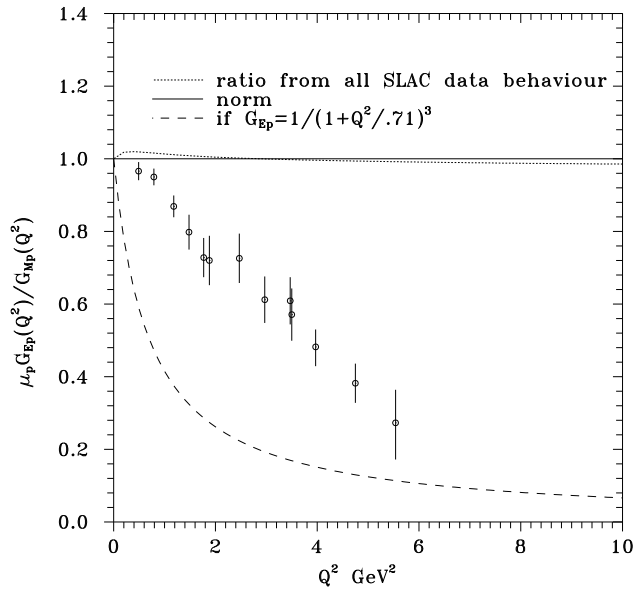}}\caption{New
JLab polarization data on the ratio $\mu_{p}G_{Ep}(t)/G_{Mp}(t)$}
%\label{decay}
\end{figure}

In order to solve this puzzle, one has to notice at the expression of the
differential cross-section of the unpolarized elastic scattering of
electrons on protons that due to the coefficient -$Q^{2}/4m_{p}^{2}$ in the
front of $G_{Mp}^{2}(Q^{2})$ the contribution of $G_{Ep}^{2}(Q^{2})$ to the $%
d\sigma /d\Omega $ with increasing $Q^{2}$ is suppressed. For instance at $%
Q^{2}=3$GeV$^{2}$ the $G_{Ep}^{2}(Q^{2})$ contributes with only 5\% and with
increased $Q^{2}$ it is even less. Then how it was possible by the
Rosenbluth technique to draw out from $d\sigma /d\Omega $ so precise data on
$G_{Ep}^{2}(Q^{2})$ almost up to $Q^{2}\approx 35$GeV$^{2}$? This question
led us to the conjecture that may be data on $G_{Ep}(Q^{2})$ in the
space-like region extracted by Rosenbluth technique are unreliable and they
should be ignored in the global analysis.

On the other hand, may be the very precise data on $\mu
_{p}G_{Ep}(Q^{2})/G_{Mp}(Q^{2})$ obtained by measuring $P_{t}$ and $P_{l}$
of recoil proton's polarization in Jefferson Lab do not contradict
predictions of pQCD (their steeper falling in comparison with the dipole
behavior one could understand as finite momentum effect) and they are not in
disagreement with all other known nucleon FF properties including also all
other existing nucleon FF data.

In this spirit a discussion \ is carried out with only one aim to conserve
as much as it is possible the one-photon exchange approximation validity in
the electron-proton EM interactions, in the framework of which $d\sigma
/d\Omega $ of the unpolarized elastic electron-proton scattering and also
expressions for transverse $P_{t}$ and longitudinal $P_{l}$ components of
the recoil proton's polarization in the electron scattering plane of the
polarization transfer process $\overrightarrow{e}^{-}p\rightarrow e^{-}%
\overrightarrow{p}$ were derived.

Because recently it was suggested [11-13] that the two-photon corrections
could resolve a large part of the discrepancy between the two abovementioned
experimental techniques in the Born approximation. And if two-photon
corrections play so important role in the elastic electron-proton
scattering, what about an importance of two photon corrections in deep
inelastic scattering (DIS), which could have unforseen consequences.

\begin{figure}[htb]
\centerline{\includegraphics[width=0.45\textwidth]{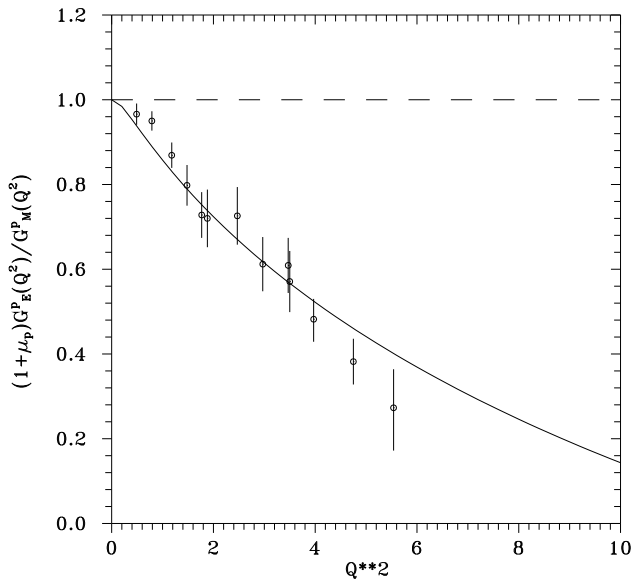}}\caption{Description
of the new JLab polarization data by ten-resonance unitary and
analytic nucleon structure model}
%\label{decay}
\end{figure}

Therefore, in order to verify our above formulated conjecture,
further we exclude all older space-like data on $G_{Ep}(t)$
obtained by Rosenbluth technique, then we substitute them only by
Jefferson Lab polarization data on $\mu
_{p}G_{Ep}(Q^{2})/G_{Mp}(Q^{2})$ and analyze the new data together
with all electric proton time-like data and all space-like and
time-like magnetic proton and electric and magnetic neutron data
in the framework of the unitary and analytic nucleon EM structure
model, formulated in Sec. 2 and comprising all nucleon FF
properties, including also the asymptotic behavior of $G_{Ep}(t)$
predicted by pQCD.

As a result parameters of the model are changed very little, a perfect
description of new JLab data is achieved (see Fig.6), almost nothing is
changed in a description of $G_{Mp}(t),$ $G_{En}(t)$ and $G_{Mn}(t)$ in
both, the space-like and time-like regions and $|G_{Ep}(t)|$ in the
time-like region (see Figs.3,4). However, new JLab data on $\mu
_{p}G_{Ep}(Q^{2})/G_{Mp}(Q^{2})$ strongly require the existence of a zero
(see full-line in Fig.3a.), i.e. the diffraction minimum in space-like
region of $|G_{Ep}(t)|$ around $t=-Q^{2}=-13$GeV$^{2},$ which could change
the charge distribution behavior within proton.

\section{Charge distribution within the proton}

The proton charge distribution (assuming to be spherically symmetric) is an
inverse Fourier transform of the proton electric FF%
\begin{equation}
\rho _{p}(r)=\frac{1}{(2\pi )^{3}}\int e^{-iQr}G_{Ep}(Q^{2})d^{3}Q
\end{equation}%
from where%
\begin{equation}
\rho _{p}(r)=\frac{4\pi }{(2\pi )^{3}}\int_{0}^{\infty }G_{Ep}(Q^{2})\frac{%
\sin (Qr)}{Qr}Q^{2}dQ.
\end{equation}

\begin{figure}[htb]
\centerline{\includegraphics[width=0.45\textwidth]{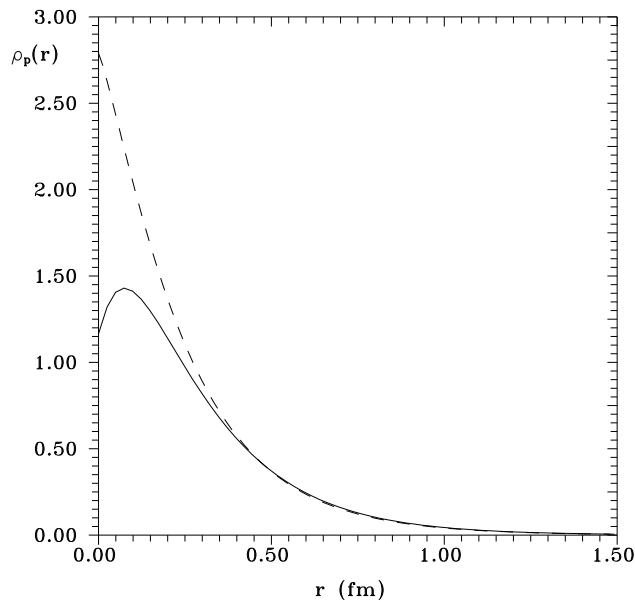}}\caption{Charge
distribution behavior within the proton}
%\label{decay}
\end{figure}

Substituting for the $G_{Ep}(Q^{2})$ under the integral either the
older behavior from Fig.3a given by dashed line, or the new
behavior with the zero (see Fig.3a given by full line) following
from the new JLab polarization data, one gets different charge
distributions within the proton given in Fig.7 by dashed and full
lines, respectively. That all leads also to different mean square
proton charge radii. The old proton charge radius takes the value
$\left\langle r_{p}^{2}\right\rangle =0.68$fm$^{2}$ and the new
one $\left\langle r_{p}^{2}\right\rangle =0.72$fm$^{2}.$

\section{Conclusions}

On the basis of analysis of the new JLab polarization data on $\mu
_{p}G_{Ep}(Q^{2})/G_{Mp}(Q^{2})$ in the framework of the ten-resonance
unitary and analytic model we came to the conclusion, that they are
consistent with all known nucleon FF properties and with all other existing
FF data as well, besides the space-like data on $G_{Ep}(Q^{2})$ obtained by
Rosenbluth technique. They do not contradict the asymptotics predicted by
pQCD, however they require an existence of zero around $t=-Q^{2}=-13$GeV$%
^{2} $ and reveal a new knowledge on the proton charge distribution, leading
to a larger mean square proton charge radius value.


\begin{thebibliography}{10}
\bibitem[1]{} P.Mergel, U.-G.Meissner and D.Drechsel, \emph{Nucl. Phys.} A
596 367 1996

\bibitem[2]{} H.-W.Hammer,U.-G.Meissner and D.Drechsel, \emph{Phys. Lett.} B
385 343 1996

\bibitem[3]{} S.Furuichi and D.Watanabe, \emph{Nuovo Cimento} A 110 577 1997

\bibitem[4]{} M.K.Jones et al, \emph{Phys. Rev. Lett.} 84 1398 2000

\bibitem[5]{} O.Gayon et al, \emph{Phys. Rev. Lett.} 88 092301-1 2002

\bibitem[6]{} S.Dubni\v{c}ka, A.Z.Dubni\v{c}kov\'{a} and P.Weisenpacher,
\emph{J. Phys.} G 29 405 2003

\bibitem[7]{} S.Dubni\v{c}ka, A.Z.Dubni\v{c}kov\'{a} and P.Weisenpacher,
\emph{Eur. Phys. J.} C 32 381 2004

\bibitem[8]{} K.Hagivara et al, \emph{Phys. Rev.} D 66 010001 2002

\bibitem[9]{} M.E.Biagini, S.Dubni\v{c}ka, E.Etim and P.Kol\'{a}r, \emph{%
Nuovo Cimento} A 104 363 1991

\bibitem[10]{} T.Frederico, H.-Ch.Pauli and Shan-Sui Zhan, \emph{Phys. Rev.}
D 66 116011 2002

\bibitem[11]{} P.A.M.Guichon and M.Vanderhaeghen, \emph{Phys. Rev. Lett.} 91
142 304-1 2003

\bibitem[12]{} P.G.Blunder, W.Melnitchouk and J.A.Tjon, \emph{Phys. Rev.
Lett.} 91 142 304-1 2003

\bibitem[13]{} Y.-C.Chen et al, \emph{Phys. Rev. Lett.} 93 122 301-1 2004
\end{thebibliography}
\end{document}